\documentclass[journal,twoside,web]{ieeecolor}
\usepackage{lcsys}
\usepackage{cite}
\usepackage{amsmath,amssymb,amsfonts}
\usepackage{graphicx}
\usepackage{textcomp}
\usepackage{booktabs}  
\usepackage{algorithm}
\usepackage{algpseudocode}
\usepackage{array}

\usepackage{xcolor}
\usepackage[colorlinks,bookmarks=false]{hyperref}
\hypersetup{citecolor=[RGB]{0,0,255}, urlcolor=[RGB]{0,0,255}}

\newtheorem{theorem}{Theorem}

\newtheorem{remark}{Remark}

\newcommand{\revise}[1]{\textcolor{black}{#1}}
\DeclareMathOperator*{\argmin}{argmin} 

\def\BibTeX{{\rm B\kern-.05em{\sc i\kern-.025em b}\kern-.08em
    T\kern-.1667em\lower.7ex\hbox{E}\kern-.125emX}}
\begin{document}

\title{Stability-Constrained Learning for Frequency Regulation in Power Grids with Variable Inertia} 

\author{Jie Feng$^{1}$, Manasa Muralidharan$^{2}$, Rodrigo Henriquez-Auba$^{3}$, Patricia Hidalgo-Gonzalez$^{1,2}$, and Yuanyuan Shi$^{1}$\vspace{-3em}
\thanks{This work is supported in part by UC-National Laboratory In-Residence Graduate Fellowship L24GF7923, NSF ECCS-2200692, and Jacobs School Early Career Faculty Development Award 2023.  }
\thanks{$^{1}$ Jie Feng, Patricia Hidalgo-Gonzalez and Yuanyuan Shi are with the Department of Electrical Engineering, UC San Diego, CA, USA. 
}
\thanks{$^{2}$ Manasa Muralidharan and Patricia Hidalgo-Gonzalez are with the Department of Mechanical and Aerospace Engineering and the Center for Energy Research, UC San Diego, CA, USA. 
}
\thanks{$^{3}$ Rodrigo Henriquez-Auba is with the National Renewable Energy Laboratory, Golden, Colorado.}} 

\maketitle

\thispagestyle{empty} 
\vspace{-20cm}
\begin{abstract}
The increasing penetration of converter-based renewable generation has resulted in faster frequency dynamics, and low and variable inertia. As a result, there is a need for frequency control methods that are able to stabilize a disturbance in the power system at timescales comparable to the fast converter dynamics. This paper proposes a combined linear and neural network controller for \revise{inverter-based} primary frequency control that is stable at time-varying levels of inertia. We model the time-variance in inertia via a switched affine hybrid system model. We derive stability certificates for the proposed controller via a quadratic candidate Lyapunov function. We test the proposed control on a 12-bus 3-area test network, and compare its performance with a base case linear controller, optimized linear controller, and \revise{finite-horizon Linear Quadratic Regulator} (LQR). Our proposed controller achieves faster mean settling time and over $50\%$ reduction in average control cost across $100$ inertia scenarios compared to the optimized linear controller. Unlike \revise{LQR} which requires complete knowledge of the inertia trajectories and system dynamics over the entire control time horizon, our proposed controller is real-time tractable, and achieves comparable performance to \revise{LQR}. 

\end{abstract}

\section{Introduction}
\IEEEPARstart{R}enewable energy has experienced a rapid growth, and is predicted to constitute one-third of the global generation mix by 2025\cite{IEA2023}. Unlike conventional generators, renewables are interfaced with the grid via power electronic inverters, which  lack rotational inertia. In synchronous generator-dominated grids, aggregated rotational inertia slows down the system frequency response in the event of an imbalance between power supply and demand  \cite{Kundur1994}. This allows controllers sufficient response time to restore frequency to its nominal value. In contrast, renewable-dominated grids are characterized by low and time-varying inertia, and fast frequency dynamics \cite{Ulbig2014}. This creates a need for real-time tractable, and fast-acting controllers (in the order of milliseconds) \cite{Ulbig2013} that are capable of stabilizing a disturbance without impacting frequency-dependent load shedding and other protection schemes \cite{Markovic2021}.

Several works have proposed virtual inertia and damping allocation as a solution to address the frequency stability challenges in low inertia systems \cite{Gross2017,Poolla2019,Venkatraman2021}. The proposed approaches vary in the order of their frequency dynamics models and in the metrics used for characterizing controller performance (see \cite{Milano2018} for a comprehensive overview). \revise{Nevertheless, it's challenging to extend the controllers designed for constant inertia to a time-varying system, given that even exponentially stable subsystems can become unstable with time-varying parameters \cite{VU2007639}.} Only a few papers have considered frequency control for variable inertia systems \cite{Hidalgo2018,Misyris2018,Guo2019,Hidalgo2019,Srivastava2022}. The work in \cite{Misyris2018} proposes a robust controller that optimizes the worst case system performance via a $\mathcal{H}_\infty$ loop shaping controller that adapts to time-varying frequency and damping in low-inertia systems. The work in \cite{Guo2019} explicitly considers the temporal variation in inertia by modeling it as multiplicative and additive noise in the linearized stochastic swing dynamics model, and characterizes controller performance via the  $\mathcal{H}_2$ system norm. The work in \cite{Hidalgo2018} models the time-varying frequency dynamics as a switched affine hybrid system, with the system switching through different modes representing different levels of inertia. Using this framework, the authors solve a receding-horizon model predictive control problem for dynamic virtual inertia placement. The work in \cite{Hidalgo2019} proposes a data-driven virtual inertia controller that learns an optimal linear control gain from the \revise{finite-horizon LQR} solution that stabilizes the switched system in all inertia modes. The work in \cite{Srivastava2022} expands on \cite{Hidalgo2018,Hidalgo2019} to prove the existence of such a stabilizing time-invariant linear controller. 

There has been a lot of recent interest in implementing machine learning to address frequency regulation challenges in the power grid (see~\cite{Chen2022} for a review). 
However, \revise{the key challenge is that standard learning techniques do not provide stability guarantees.}
Particularly, a neural network controller with a low training loss may actually lead to system instabilities (i.e., unbounded frequency deviations) when implemented during testing as observed in \cite{Cui2022, Shi2022}. Stability is critical for power system operation because it can lead to catastrophic consequences, e.g., blackouts \cite{Haes2019}. To take stability into account for a nonlinear time-varying system with uncertainty, \cite{Ming2020} proposes a quadratic constraint based on restricting the partial gradients of the control policy to a bounded (safety) set to guarantee global asymptotic stability. 
Another class of papers are based on integrating Lyapunov stability constraints into neural-network-based design~\cite{Shi2022,Jiang2022,Huang2022,Feng2023,Cui2023, Feng2023l, yuan2023learning, yuan2023constraints,cui2024structured}, and enforce strict monotonicity of the policy for stability\cite{Shi2022,Jiang2022,Feng2023,Cui2023, yuan2023learning, yuan2023constraints,cui2024structured}. In particular, \cite{Jiang2022,Cui2023,cui2024structured} propose a nonlinear controller for primary frequency control in lossless power networks considering nonlinear frequency dynamics. For the proposed candidate Lyapunov function, local asymptotic stability is guaranteed as long as the controller is a monotonically increasing function that passes through the origin.
The works \cite{Shi2022,Feng2023,Feng2023l, yuan2023learning, yuan2023constraints} propose nonlinear controllers for voltage control considering the linearized distribution grid power flow model, and show that an incremental control with monotonically decreasing instantaneous control functions guarantee stability. 
For problems where it is hard to analytically derive a Lyapunov function, \cite{Cui2022, Huang2022,Chang2022} propose methods for learning a candidate Lyapunov function jointly with a controller, with both parameterized as neural networks. To the best of the authors' knowledge, no existing work has explored learning-based frequency control that guarantees stability under \emph{time-varying} frequency dynamics.

This paper leverages the power of deep learning to optimize the performance of \revise{inverter-based} frequency control under time-varying inertia while maintaining stability guarantees. 
The  contributions of this work are three-fold: 
\begin{enumerate}
 \item  a fast-acting, data-driven controller that considers frequency dynamics with time-varying inertia,
 \item that satisfies Lyapunov stability conditions for a switched affine hybrid system (i.e., in all inertia modes and switching sequences), and
 \item that performs comparably to \revise{finite-horizon LQR} in stabilizing a frequency disturbance in milliseconds (real-time computationally tractable).
 \end{enumerate}

The remainder of the paper is organized as follows. In Section \ref{sec:formulation}, we introduce the switched affine hybrid system model of frequency dynamics for a system with time-varying inertia, and the \revise{finite-horizon LQR} formulation for frequency control. In Section \ref{sec:method}, we expand on the notion of Lyapunov stability for a switched affine hybrid system in all inertia modes, and present our proposed controller. We derive Lyapunov stability constraints for the proposed controller, and propose a method to integrate the derived stability constraints into a learning algorithm. We implement the proposed stability-constrained learning algorithm on a modified Kundur 12-bus 3-area test system and present our simulation's results in Section \ref{sec:simulations}. Lastly, in Section \ref{sec:conclusion} we conclude and outline future work.

\section{Model and Problem Formulation}
\label{sec:formulation}
\subsection{Frequency Dynamics as a Hybrid Switching System}
Consider a power network described by an undirected graph, $\mathcal{G} =\{\mathcal{N},\mathcal{E}\}$, where $\mathcal{N} = \{1, \ldots, n\}$ represents the buses (nodes) of the power network and  $\mathcal{E}$ represents the edges (transmission lines). 
Assuming identical unit voltage magnitudes, and purely inductive lines, a small-signal approximation of the swing equation \cite{Poolla2017} gives us the following linearized dynamics, $\forall i\in \mathcal{N}$:
\begin{equation}\label{eq:freq_dyn_static_onenode}
    m_i\Ddot{\theta}_i(t) + d_i\dot \theta_i(t) = p_{in,i}(t) - \sum_{j\in \mathcal{N}} b_{ij}(\theta_i(t)-\theta_j(t))\revise{,}
\end{equation}
where $\theta_i$ is the voltage phase angle at node $i$, $m_i$ is the inertia coefficient at node $i$, $d_i$ is a lumped parameter representing the droop control or frequency damping coefficient at node $i$, $p_{in,i}$ represents the power input at node $i$, and $b_{ij}$ is the susceptance of the transmission line between nodes $i$ and $j$. 

\revise{Shifts towards greater integration of renewable energy sources introduce variability in the system's inertia. This variability is attributed to the varying ratios of power generation from renewables (and their associated power electronic converters) and conventional generators across different times of the day. We use  \cite{Hidalgo2019} to model these new dynamics as a switched affine hybrid system to capture the time-variance of the system and put the frequency dynamics in \eqref{eq:freq_dyn_static_onenode} in the compact vector form:
\begin{equation}\label{eq:freq_dyn}
    \underbrace{\begin{bmatrix}
    \dot{\theta}(t)\\\dot{\omega} (t)
    \end{bmatrix}}_{\dot{x}(t)}=
    \underbrace{\begin{bmatrix}
    0 & I \\
    -M_{q(t)}^{-1}L & -M_{q(t)}^{-1}D
    \end{bmatrix}}_{A_{q(t)}}
    \underbrace{\begin{bmatrix}
    \theta(t)\\\omega(t)
    \end{bmatrix}}_{x(t)}
    + \underbrace{\begin{bmatrix}
    0\\M_{q(t)}^{-1}
    \end{bmatrix}}_{B_{q(t)}} \underbrace{p_{in}(t)}_{u(t)}\revise{,}
\end{equation}
where $\theta(t) \in \mathbb{R}^n$ and $\omega(t) \in \mathbb{R}^n$ respectively denote the angle and frequency deviations from their nominal values at all nodes at time t, $x\in\mathbb{R}^{2n}$ corresponds to the stacked state vector of angle and frequency deviations at all nodes at time $t$, $u = p_{in}(t) \in \mathbb{R}^{n}$ is the control action at time $t$, $D=\text{diag}(d_i)\in \mathbb{R}^{n\times n}$ is a diagonal matrix containing the droop/damping coefficients at all nodes, and $L\in \mathbb{R}^{n\times n}$ is the Laplacian of the grid. Further, $M_{q}$ represents the inertia matrix in mode $q\in\{1,...,p\}$ and each $M_{q}=\text{diag}(m_{i, q})\in \mathbb{R}^{n\times n}$ is a diagonal matrix with $m_{i, q}$ denoting inertia constant at node $i$. Each operational mode $q$ in \eqref{eq:freq_dyn} corresponds to a specific inertia value $h_{q}$ (in seconds), based on the mix of online generators and converters at time $t$. In particular, inertia of $<2~s$ represents a renewable-dominated node, $2-4~s$ represents a hydroelectric generation-dominated node, and $4-10~s$ represents a thermal generation-dominated node \cite{Kundur1994}. The inertia coefficients are calculated as $m_{q,i} = \frac{2h_qS_{rated,i}}{\omega_{s}}$, where $\omega_{s} = 50$~Hz and $S_{rated,i}$ is the net power rating at node $i$. Note that $x$, $u$, and $q$ are all time-dependent variables; we omit the $t$ in their notation for brevity.}

\subsection{Training Set Generation} 
To train a policy that minimizes frequency deviations and control costs, we generate a training set for the learning-based controller with the following finite-horizon \revise{LQR}: 
\begin{subequations}\label{eq:MPC}
    \begin{align}
        \min_{u,x} &\quad \int_{t=0}^T \left( x^\top Qx + u^\top Ru \right) dt\revise{,} \label{eq:cost} \\
        \text{s.t.} &\quad \dot{x} = A_{q} x + B_{q} u \quad \forall t\in[0,T]\revise{,}\label{eq:state_eq} \\
        &\quad x(0)= x^{(0)}\revise{,}
    \end{align}
\end{subequations}
\noindent where $x^{(0)}$ is the initial state, states $x$ and control actions $u$ are the decision variables. The objective function includes quadratic costs for frequency deviation and control action over the optimization time horizon, where \(Q \succeq 0\) and \(R \succ 0\) can be tuned according to the desired control objectives. The \revise{finite-horizon LQR} has complete knowledge of the dynamics, including the modes, over the entire time horizon. Problem \eqref{eq:MPC} is a Quadratic Programming problem solvable with CVX~\cite{cvx}.

\section{Methodology}
\label{sec:method}
In this section, we begin by deriving a Lyapunov function for the swing dynamics with time-varying inertia described in  
\eqref{eq:freq_dyn}. Then, we derive the stability condition under this Lyapunov function and present our proposed controller. 
Finally, we introduce the training method for the proposed algorithm. Figure \ref{fig:diag} shows a diagram of the proposed policy.

\subsection{Candidate Lyapunov Function for the Switched System}\label{sec:feasible}
The first-stage objective is to find a Lyapunov function for the switched affine hybrid system. We propose to \emph{jointly} identify a Lyapunov function $V(x) = x^\top P x$ and a \emph{linear} feedback controller $u = K x$ that can stabilize all the modes. In the next part, we will show how to improve the performance of the linear controller via a neural network residual policy while maintaining the stability guarantee. 

{Formally, we aim to find a controller $u = K x$ and a common Lyapunov function $V(x) = x^\top P x, P \succ 0$ such that the following Lyapunov stability condition is satisfied,
\begin{equation}
    (A_q \!+\! B_q K)^\top P \!+\! P(A_q \!+\! B_q K) \prec 0\,,\quad \forall  q \in \{1, 2, ..., p\}, \label{eq:Lie_derivative1} 
\end{equation}
Given that the Lyapunov function $V(x) = x^\top P x$ is positive definite since $P \succ 0$ and satisfies the Lie derivative condition \eqref{eq:Lie_derivative1}, the linear feedback controller $u = Kx$ guarantees the stability of the switched affine hybrid system.}

However, a notable challenge is that the constraint \eqref{eq:Lie_derivative1}  is \emph{nonconvex} in the decision variables $K$ and $P$ due to the products of bi-linear terms. Thus, we {adopt} a change of variables following \cite[Ch. 7]{boyd1994linear}.
Defining $X = P^{-1}$ and $Y = KX$, {we formulate the joint Lyapunov function and stable control identification feasibility problem as follows,}
\begin{subequations}\label{eq:find_common_Lyapunov2}
    \begin{align}
        \min_{X, Y} &~  C \,, \\
        \text{s.t.} &~ A_q X \!+\! X \!A_q^\top \!+\! B_q Y \!+\! Y^\top\! B_q^\top \!\prec 0, \forall  q\! \in\!\{1,...,p\} ,\\
        &~ X \succ 0, \label{eq:Lie_derivative2_2}
    \end{align}
\end{subequations}
{where $C$ is a constant.} This problem is convex and feasible as \cite{Srivastava2022} analytically guarantees it.
Suppose $(X_s, Y_s)$ are solutions of \eqref{eq:find_common_Lyapunov2}. Then the feasible control gain $K_s$ is derived as $K_s = Y_s(X_s)^{-1}$ and the Lyapunov function is defined as \begin{equation}\label{common_Lyapunov}
    V = x^\top P_s x\,, \quad P_s = (X_s)^{-1}.
\end{equation} 
Therefore, no matter how many different modes the switched affine system has, we can always find a linear controller and a common Lyapunov function \cite{Srivastava2022}. 

\subsection{Proposed Controller}
\begin{figure}[t]
    \centering
    \includegraphics[width=\linewidth]{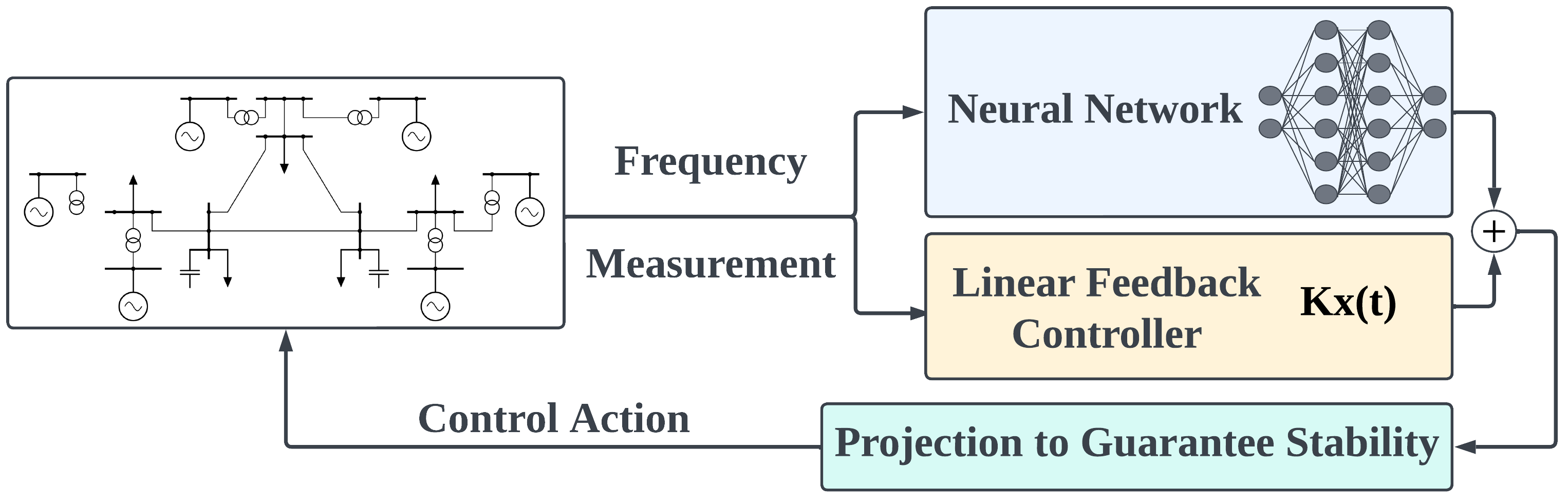}
    \vspace{-0.5cm}
    \caption{The diagram of the proposed controller depicts a combination of a linear controller $Kx(t)$ and a neural network residual $\pi_{\psi}(\cdot)$, constrained to actions that satisfy the Lyapunov stability conditions specified in \eqref{eq:closed-form}.}
    \label{fig:diag}
\end{figure}

Having identified a stable linear controller for the hybrid system and the corresponding Lyapunov function, we now introduce our proposed algorithm. To guarantee stability and improve performance, we parameterize the proposed controller as a \emph{combination} of a stabilizing linear feedback controller and a nonlinear residual $\pi_{\psi}(\cdot):\mathbb{R}^{2n}\to \mathbb{R}^{n}$ (Fig~\ref{fig:diag}). The nonlinear residual is parameterized as a neural network with parameter $\psi$, and its inputs are the current states $x$. We consider a controller defined as follows: 
\begin{equation}\label{eq:controller}
    u = Kx + \pi_{\psi}(x){.}
\end{equation}
By combining the linear controller with a neural network residual, we have a dual benefit: it can inherit the guarantee of the linear control while gaining additional flexibility for performance optimization with a neural network. 

To ensure the stability of the closed-loop system we define $A_{cl,q}=A_q + B_q K$ and we utilize the Lyapunov function from \eqref{common_Lyapunov} to derive the algebraic stability constraint, $\forall x \neq 0$: 
\begin{align} \label{eq:stability_condition_proposed_controller}
    \dot{V} & =x^\top \left(A_{cl,q}^\top P+P A_{cl,q}\right)x 
 +2x^\top PB_{q} \pi_{\psi}(x) < 0 \,.
\end{align}

We introduce short-handed notations $V_{\text{Lin}}(x,q): = x^\top \left(A_{cl,q}^\top P+P A_{cl,q}\right)x$ 
and $g(x,q)=2 B_{q}^\top P x$. Thus, \eqref{eq:stability_condition_proposed_controller} can be compactly written as,
\begin{align}\label{eq:stability_condition_proposed_controller2} 
    V_{\text{Lin}}(x,q) + g^\top(x,q) \pi_{\psi}(x) < 0 \,, \forall x \neq 0 {.}
\end{align}
The switching system is asymptotically stable with the proposed policy if \eqref{eq:stability_condition_proposed_controller2} holds for all time $t$. {We summarize the main result as the following Theorem.
\begin{theorem}
        Consider a controller defined as $u_{\psi}(x) = Kx + \Pi[\pi_{\psi}(x)]$, where $u=Kx$ is a linear controller that can stabilize the switching system \eqref{eq:freq_dyn}, and $\pi_{\psi}(x)$ is a neural network controller parameterized by $\psi$. For all $x \neq 0$, define the projection operation $\Pi[\pi_{\psi}(x)]$ as follows \eqref{eq:closed-form},
        \begin{subequations}\label{eq:closed-form}
        \begin{align}
            &\lambda^* = \left [ \frac{g^\top(x,q)\pi_{\psi}(x) +V_{\text{Lin}}(x,q)+\epsilon}{g^\top(x,q)g(x,q)}\right]^+,\label{eq:closeda}\\
            & \Pi[\pi_{\psi}(x)] = \pi_{\psi}(x) - \lambda^*g(x,q),\label{eq:closedb}
        \end{align}
        \end{subequations}
       where $\epsilon>0$ is a sufficiently small constant. For $x=0$, the projection is $\Pi[\pi_{\psi}(0)] = 0$. Then the closed-loop system \eqref{eq:freq_dyn} is asymptotically stable with respect to the origin for arbitrary switching signal $q(t):\mathbb{R}^+ \mapsto \{1, ..., p\}$. 
\end{theorem}
\begin{proof}
Consider the following convex programming, 
    \begin{subequations}\label{eq:stability_layer}
    \begin{align}
        &\Pi[\pi_{\psi}(x)] = \argmin_{\xi} \frac{1}{2}\lVert \xi - \pi_{\psi}(x)\rVert^2,\\ \label{eq:constraint}
        &\text{s.t. }  V_{\text{Lin}}(x, q) + g^\top(x,q) \xi \leq -\epsilon.
    \end{align}
    \end{subequations}
     Given that $K$ is a stable linear controller for the system, with the common Lyapunov function $V = x^\top P x$, thus for any $q \in \{1, ..., p\}$
   $V_{\text{Lin}}(x, q) <0\,, \forall x\neq 0.$ Therefore, the project problem~\eqref{eq:stability_layer} is always feasible with a solution $\xi=0$. Consider the Lagrangian of \eqref{eq:stability_layer}
    \begin{align}
        \mathbb{L}(\xi,\lambda)&=\frac{1}{2}\lVert \xi-\pi_{\psi}(x)\rVert^2
    +\lambda \left(V_{\text{Lin}}(x,q)+g^\top(x,q)\xi+\epsilon\right),\nonumber
    \end{align}
    The optimal primal and dual solutions $(\xi^*,\lambda^*)$ satisfy the {Karush-Kuhn-Tucker (KKT)} conditions, 
\begin{subequations}
\begin{align}
    &\nabla_\xi \mathbb{L} =\xi^*-\pi_{\psi}(x)+\lambda^*g(x,q)=0,\label{eq:KKT1}\\
    &\lambda^*(V_{\text{Lin}}(x,q)+g^\top(x,q)\xi^*+\epsilon)=0\,, \label{eq:KKT2}\\
    &{V_{\text{Lin}}(x, q) + g^\top(x,q) \xi \leq -\epsilon, {\lambda^*\geq 0}.}
\end{align}
\end{subequations}
Thus, \eqref{eq:closedb} is a direct result from \eqref{eq:KKT1}. Substituting \eqref{eq:closedb} in \eqref{eq:KKT2} gives \eqref{eq:closeda}. As a result, \eqref{eq:closed-form} provides an optimal solution of \eqref{eq:stability_layer}.
This solution guarantees that $V_{\text{Lin}}(x,q) + g^\top(x,q) \Pi[\pi_{\psi}(x)] < 0 \,, \forall x \neq 0$ holds for all time $t$, which leads to asymptotic stability with respect to the origin.
\end{proof}
}

Therefore, the final designed policy reads as follows, 
\begin{equation}\label{eq:controller_final}
    u_{\psi}(x) = Kx + \Pi[\pi_{\psi}(x)]\,,
\end{equation}
where $\Pi[\pi_{\psi}(x)]$ is given by \eqref{eq:closed-form}.
{\begin{remark}
    To analyze the impact of linearization error on stability, consider the following nonlinear dynamics,
    $$\dot{x}=A_qx+B_qu+\eta(x),$$
    where $\eta(x)$ is the linearization error depending on state. With the same Lyapunov function $V=x^\top Px$, the stability condition for the nonlinear dynamics reads as follows, 
\begin{align*} \label{eq:stability_condition_proposed_controller}
    \dot{V} & =\underbrace{x^\top \left(A_{cl,q}^\top P+P A_{cl,q}\right)x 
 +2x^\top PB_{q}\pi_{\psi}(x)}_{V_{\text{Lin}}(x, q) + g^\top(x,q) \xi \leq -\epsilon} +2x^\top P \eta(x)\,. 
\end{align*}
Therefore, our stability constraint in \eqref{eq:constraint} guarantees stability of the nonlinear dynamics in a local region around the origin when the linearization error is small.
\end{remark}}

\subsection{Stability-Constrained Learning for Frequency Regulation}
{In Section \ref{sec:feasible}, we introduced feasible solutions for the feedback control gain $K$ and the Lyapunov function obtained by solving \eqref{eq:find_common_Lyapunov2}. However, the direct solution of \eqref{eq:find_common_Lyapunov2} often leads to suboptimal outcomes. Indeed, it has been observed in previous studies~\cite{Cui2023} that neural network-based controllers can reduce the control cost by 30\% and shorten the frequency recovery time by a third compared to a linear control policy. 
To address this, we propose an iterative approach in this section to jointly optimize the neural network residual controller, the linear feedback control gain, and the Lyapunov function.} The proposed training method bypasses the need for hand-tuning, as for the BMI algorithm \cite{Srivastava2022} or the feature selection required for a regression-based linear controller~\cite{Hidalgo2019}. 

With the training set $(x_{{lqr}},u_{{lqr}})$ generated through {finite-horizon LQR} by solving \eqref{eq:MPC}, we optimize the proposed controller to mimic the behavior of the {LQR} controller. Considering the distinct parameterization of the linear controller and the neural network, 
we decompose the optimization of the proposed controller into two sub-problems: (1) optimizing the nonlinear residual $\pi_{\psi}(x)$ and (2) optimizing the linear feedback controller $Kx$ and the Lyapunov stability certificate $V(x)$. The optimization problem for $\psi$ is
\begin{equation} \label{eq:L_nonlinear}
    \min_{\psi} \quad \mathfrak{L}_\psi = \lVert u_{\psi}(x_{{lqr}}) - u_{{lqr}}\rVert{,}
\end{equation}
where $u_{\psi}$ is defined in \eqref{eq:controller_final} and $\lVert \cdot\rVert $ is the Frobenius norm. 

Then, we fix $\psi$ and optimize the linear controller and Lyapunov function. This sub-problem is formulated as an optimization problem with Lyapunov stability constraints. We use the Cholesky decomposition{\cite{cholesky}} of $P$ to enforce the positive definiteness of the Lyapunov function, i.e. $P=QQ^\top$, where $Q$ is a lower triangular matrix with positive diagonal entries. The stability constraint \eqref{eq:Lie_derivative1} is formulated as a soft penalty and verified post-training. We parameterize the Cholesky decomposition matrix $Q$ and the linear controller gain matrix $K$ as the learnable parameters. 
\begin{subequations}\label{eq:L_linear}
    \begin{align}
    &\min_{K,Q}  \mathfrak{L}_{(K,Q)} =c_1\lVert K\rVert + c_2 \lVert Kx_{{lqr}}\rVert +\lVert u_\psi(x_{{lqr}}) - u_{{lqr}}\rVert \nonumber\\
         & +c_3\sum_{i=0}^{2n}\sum_q\max(0,\text{eig}_i(A_{cl,q}^\top P + P A_{cl,q})){,}\nonumber\\
         & \quad P = QQ^\top, \text{ } Q \text{ is lower triangular,}\\
         &\quad Q_{ii}\in\mathbb{R}, Q_{ii}>0,  \forall i \in\{1,...,n\},
\end{align}
\end{subequations}
where $\psi$ is fixed, only $K$ and $Q$ are optimized. The coefficients $c_1,c_2,c_3$ are objective weights.
The second row for $\mathfrak{L}_{(K,Q)}$ is a summation of all positive eigenvalues of $A_{cl,q}^\top P + PA_{cl,q} \prec 0\,, \forall  q \in\{1,...,p\}$ to penalize violations of the Lyapunov stability constraint \eqref{eq:Lie_derivative1}. 
We deploy a warm start using the feasible solution of \eqref{eq:find_common_Lyapunov2} for $K$ and $Q$, where $Q$ is decomposed from $P_s$, such that we start with a stabilizing but suboptimal solution, improving learning efficiency. 
We minimize the norm of the control gain $\lVert K\rVert$ and the linear control action $\lVert Kx_{{lqr}}\rVert$ to avoid the large overshoot observed in the feasible solution. With $\pi_\psi(x)=0$, the same procedure solely optimizes the linear controller with its Lyapunov function, termed \emph{Linear-opt}.

The nonlinear and linear policies are updated iteratively using gradient descent {until convergence or the maximum step limit is reached, as outlined in Algorithm \ref{alg:satrl}. 
It's worth noting that while this iterative algorithm does not guarantee global optimality, the numerical experiments show that the trained controller \eqref{eq:controller_final} achieves performance comparable to the {LQR} controller with stability guarantees.}

\begin{algorithm}[t]
    {\footnotesize{
	\caption{Stability-constrained Learning}
	\label{alg:satrl}
	\begin{algorithmic}[1]
	\Ensure NN $\pi_\psi(x)$; $K,Q$; dataset $(x_{{lqr}},u_{{lqr}})$; feasible solution $K_s, P_s$;  epoch number $N_{ep}$, batch number $N_b$, batch size $N_s$, training step for linear controller $N_l$; constants $c_{1}, c_2, c_3$, learning rates $\eta_1$, $\eta_2$. 
        \State{Initialize $K,Q$ with $K_s, P_s$;
        }
	\For {$i = 0$ to $N_{ep}$}
	  \For{$j = 0$ to $N_b$}
        \State{Randomly sample $N_s$ pairs from $(x_{{lqr}},u_{{lqr}})$;}
        \State{Update $\psi$ by $\psi = \psi-\eta_1\nabla_\psi \mathfrak{L}_\psi$ {in \eqref{eq:L_nonlinear}};}
        \For{$\text{step}=0$ to $N_l$}
        \State{Update: $K=K-\eta_2\nabla_K \mathfrak{L}_{(K,Q)}$, $Q=Q-\eta_2\nabla_Q \mathfrak{L}_{(K,Q)}$ {in \eqref{eq:L_linear}}.}
        \EndFor
        \EndFor
	\EndFor
	\end{algorithmic}
 }}
\end{algorithm}

\section{Simulations}
\label{sec:simulations}
In this section, we demonstrate the effectiveness of the proposed algorithm via numerical experiments. We compare against the {LQR}, the linear controller, i.e., (\emph{Linear}), and the optimized linear controller, i.e., (\emph{Linear-opt}).

\subsection{Experimental Setup}
\begin{figure}[t]
    \centering 
    \includegraphics[width=0.85\linewidth]{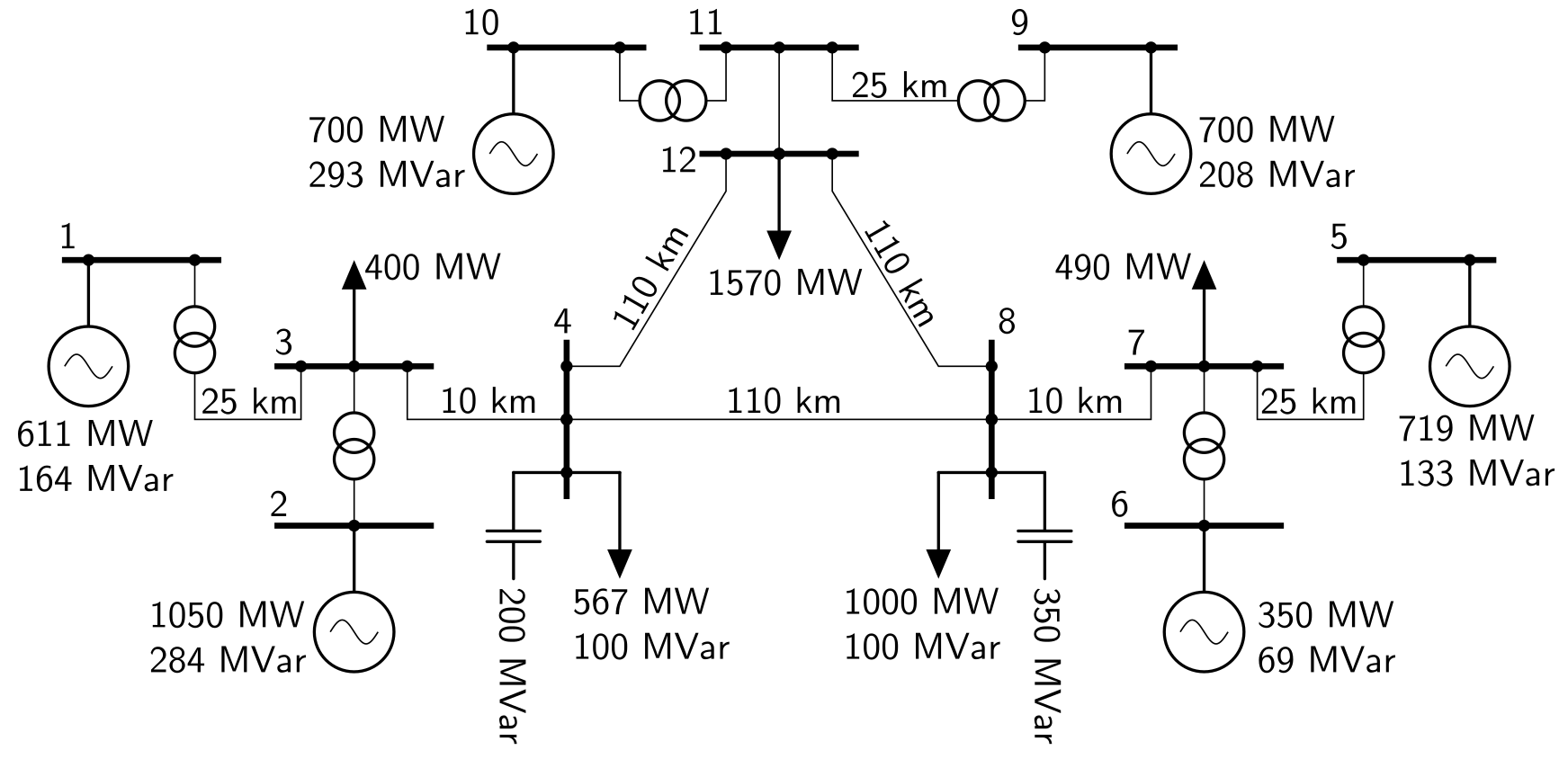}
    \vspace{-0.5cm}
    \caption{230kV/100 MVA Kundur 12-bus 3-area test network with 0.0001 + 0.001 p.u. line impedance. \vspace{-0.8cm}}
    \label{fig:network}
\end{figure}

We implement the proposed controller in the modified Kundur 12-bus 3-region network \cite{Borsche2015} (cf. Fig.~\ref{fig:network}), with discretized network dynamics using zero-order hold. 
{The inverter-based generation resources are deployed for frequency control.} We consider 9 inertia modes, with the inertia in each mode $h_q \in \{0.5, 1, 1.5, 2, 2.5, 3, 3.5, 5, 9\}$~s. For simplicity, we assume the same normalized inertia coefficient for all nodes in a given mode $q$, i.e., $m_{q,i} = m_q = h_q, \quad \forall i$.  
 The droop coefficient is 0.5 at all nodes and the {control} frequency is 1 kHz. We solve the {finite-horizon LQR} problem \eqref{eq:MPC} using \(Q = \text{diag}(0,5e^4I)\) and \(R = I\) {and collect 200 trajectories}, assuming both frequency and actions in per unit (p.u.). {The nonlinear residual is modeled using a three-layer fully connected neural network with 300 and 400 hidden units, respectively. The network takes the state $x$ as input and outputs the control actions (real power set-points).}

{Online inertia modes are estimated by a neural network pre-trained with the collected trajectories and used by \eqref{eq:closed-form} for the projection, where the input to the neural network is two consecutive frequency observations $\omega(t),\omega(t+1)$ and the corresponding action $u(t)$. The estimator achieves $90.64\%$ accuracy for inertia mode classification.} We optimize the proposed controller {by Algorithm \ref{alg:satrl} with} $c_1=0.1, c_2=0.01, c_3=500, \eta_1=0.001, \eta_2=0.01, N_{ep}=N_{b}=300, N_s=256, N_l=5$, and compare its performance against {LQR}, \emph{Linear} and \emph{Linear-opt} in 100 distinct frequency deviation scenarios, each with a time horizon of 1~s (1000 steps) and a random initial frequency deviation at each bus sampled from a uniform distribution $\mathcal{U}_{[-0.3, 0.3]}$~Hz. 
The scenarios commence in random operational modes \(q(0) \in \{1, \ldots, 9\}\). Subsequently, based on a uniform distribution, the inertia of the system either remains constant, increases or decreases every 0.1~s.

 We utilize three key metrics to evaluate our controller: (i) \textbf{settling time}, defined as the average duration for the controller to reduce frequency deviations to under 0.01 Hz; (ii) \textbf{overshoot}, which is the average maximum frequency deviation observed in each scenario; and (iii) \textbf{average cost} (Avg Cost) $\frac{1}{N}\sum_{i=0}^{N=100}\sum_{t=0}^{T=1000} x_{i}^\top Qx_{i} + u_{i}^\top Ru_{i}$.
\subsection{Results}
In 100 scenarios (cf. Table \ref{tab:performance}), the proposed controller matches {LQR}'s performance and outperforms \emph{Linear} and \emph{Linear-opt} in settling time and control cost due to the flexibility of the nonlinear residual.

\begin{figure}[h]
    \centering
    \includegraphics[width=0.95\linewidth]{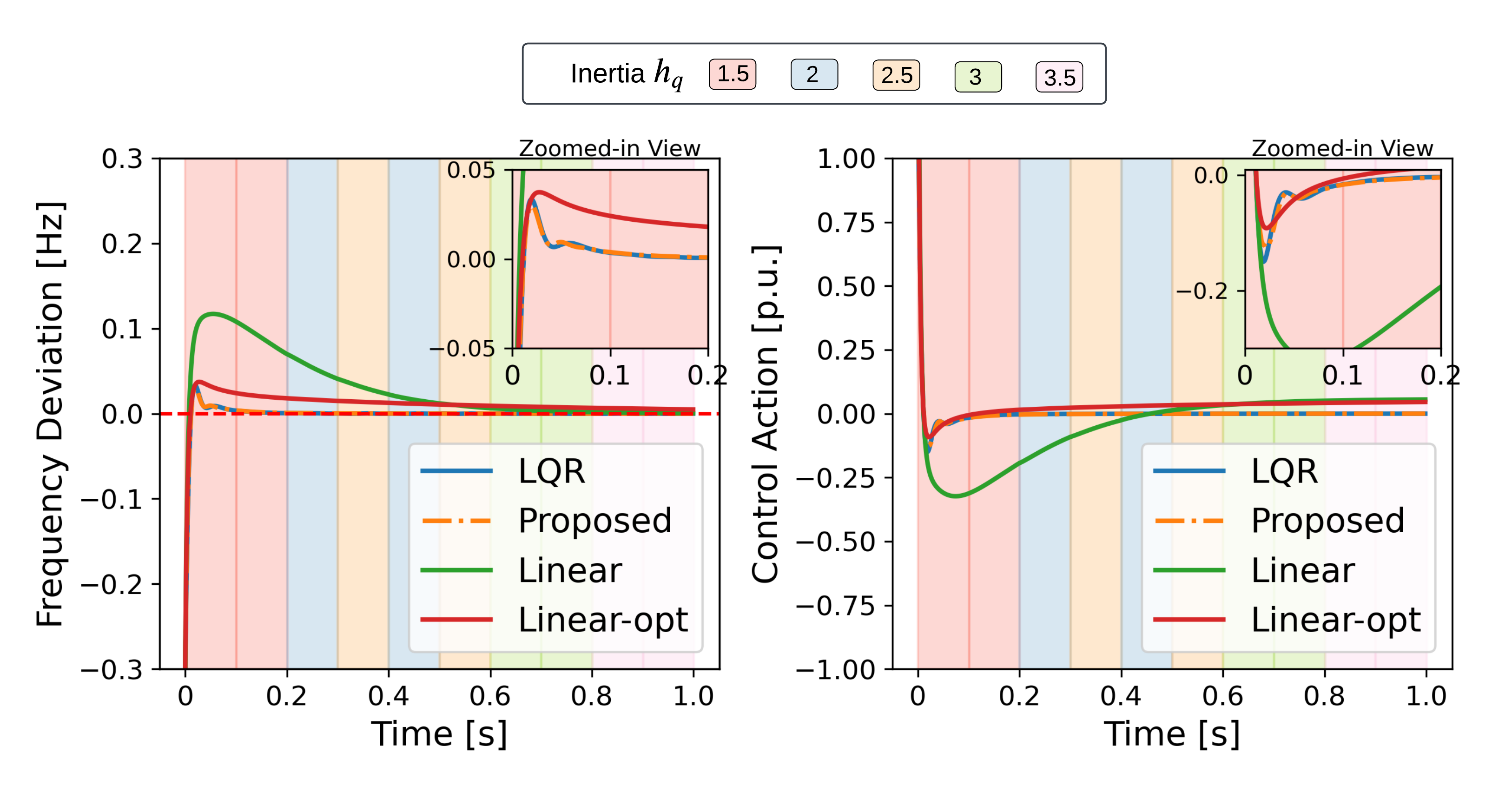}
    \vspace{-0.5cm}
    \caption{State and control trajectories for {LQR}, the proposed controller, \emph{Linear}, and \emph{Linear-opt} at bus 9 with a zoomed-in view for 0-0.2~s. The background color represents the inertia modes.\vspace{-0.5cm}}
    \label{fig:onebus}
\end{figure}

\begin{figure}[h]
    \centering
    \includegraphics[width=0.95\linewidth]{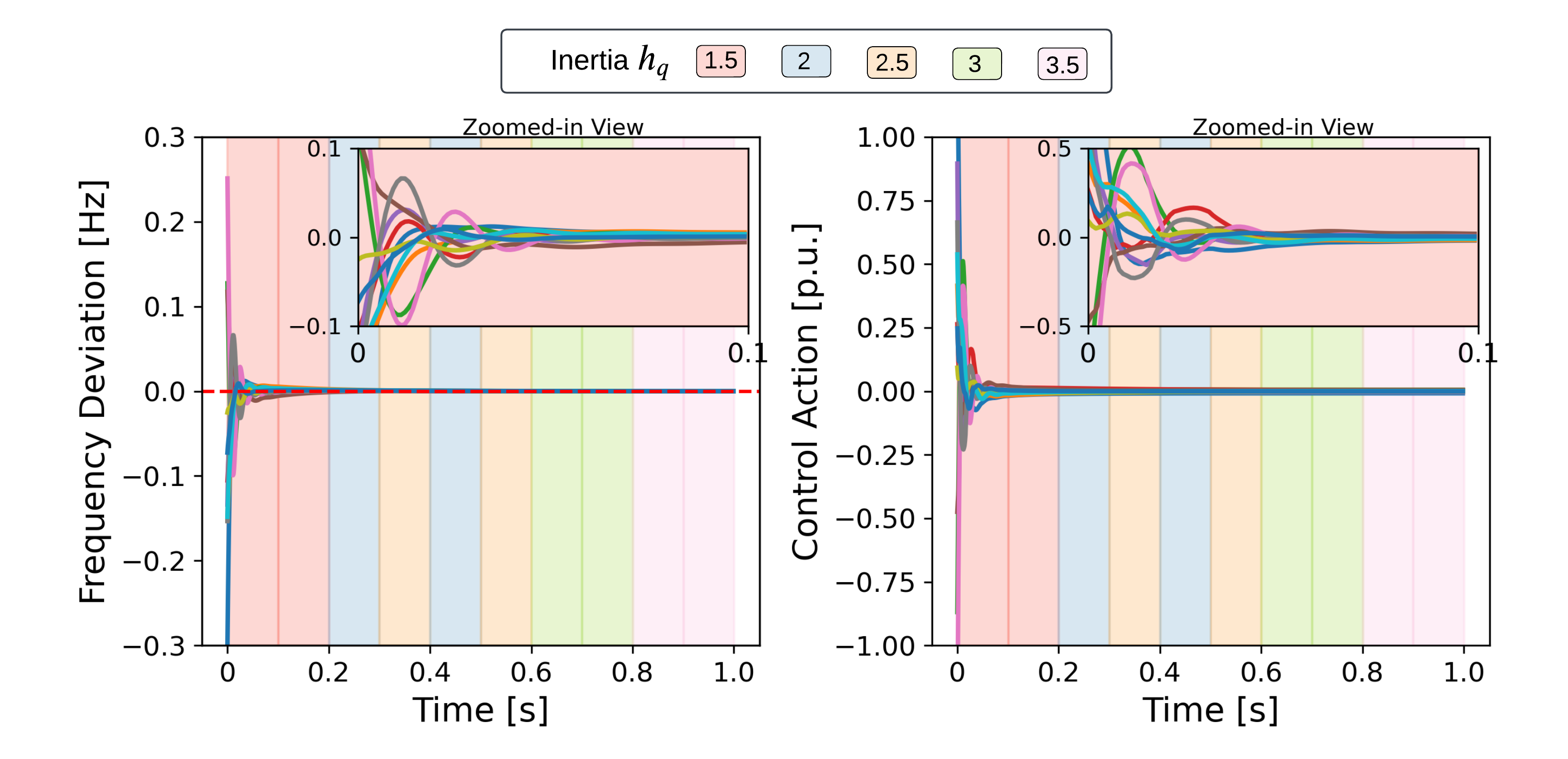}
    \vspace{-0.5cm}
    \caption{State and control trajectories of all buses with the proposed controller with a zoomed-in view for 0-0.1~s.\vspace{-0.5cm}}
    \label{fig:allbus}
\end{figure}

\begin{table}[htb!]
    \centering
    \caption{Performance of {LQR},  base case (Linear), optimized linear (Linear-opt), and proposed controller across 100 scenarios.}
    \begin{tabular}{c|cccccc}
    \toprule
    & \multicolumn{2}{c}{Settling time (ms)}  & \multicolumn{2}{c}{Overshoot (Hz) } & \multicolumn{2}{c}{Avg Cost}  \\
         Method & Mean & Std & Mean & Std & Mean & Std \\
         \midrule
        {LQR} & 122.8&77.1& {0.097}&{0.033}& 99.0&76.18\\
        \textbf{Proposed} & 147.7 &84.8& {0.096} & {0.035}& 107.1&82.39\\
        Linear & 554.8&24.2&{0.269}&{0.021}&4497.4&130.5\\
         Linear-opt & 605.2&48.8& {0.083}&{0.034}& 244.7&92.97\\
        \bottomrule
    \end{tabular}
    \label{tab:performance}
\end{table} 
Figures \ref{fig:onebus} and \ref{fig:allbus} illustrate state and control trajectories for one inertia switching scenario, where bus 9 has an initial frequency deviation of -0.3 Hz and other buses have random deviations. From Figure \ref{fig:onebus}, all controllers stabilize bus 9 despite the inertia change. Specifically, \emph{Linear} induces a relatively large overshoot and large control actions, while \emph{Linear-opt} reduces these but has a longer frequency restoration time ($\approx1$~s).  Moreover, the control action of both linear controllers might fail to converge to the optimal solution once the frequency is restored. The proposed controller, similar to {LQR}, achieves fast frequency recovery within $0.1$~s and low cost across all buses (cf. Figures \ref{fig:onebus} and \ref{fig:allbus}), demonstrating the efficiency of our algorithm.

\subsection{Controller Constraints}
We explore hardware constraints and nonlinear power flow dynamics' impact. Integrating action constraints, $\underline{u} \leq u \leq \bar{u}$, into both {LQR} formulation and projection \eqref{eq:closed-form}, we train the controller with {LQR} solutions that incorporate these constraints. Table \ref{tab:performance_cons} outlines the constrained controller's performance, where $|u_i| \leq 0.5$ p.u. for all $i \in \mathcal{N}$. Despite a slight increase in control cost and settling time compared to {LQR}, our controller still achieves fast frequency recovery.
\begin{table}[htb!]
    \centering
    \caption{{Performance of {LQR} ({LQR}-c) and proposed controller (Proposed-c) with hardware constraints.}}
    \begin{tabular}{c|cccccc}
    \toprule
    & \multicolumn{2}{c}{Settling time (ms)}  & \multicolumn{2}{c}{Overshoot (Hz) } & \multicolumn{2}{c}{Avg Cost}  \\
         Method & Mean & Std & Mean & Std & Mean & Std \\
         \midrule
        \textbf{{LQR}-C} & 143.7 &84.7& 0.114 & 0.039& 106.2&81.1\\
        \textbf{Proposed-C} & 152.2 &79.1& 0.124 & 0.043& 141.7&112.3\\
        \bottomrule
    \end{tabular}
    \label{tab:performance_cons}
\end{table}

\section{Conclusion}\label{sec:conclusion}
We propose a stability-constrained data-driven controller for frequency regulation with time-varying inertia. Our method integrates a linear controller with a neural network-based nonlinear residual, where the linear controller can stabilize the switching system with a joint Lyapunov function. During training, the linear controller, the corresponding Lyapunov function, and the nonlinear residual are optimized iteratively in an end-to-end manner. The stability of the closed-loop system is further enforced by projecting the nonlinear residual to guarantee the Lyapunov condition. Thanks to the nonlinear residual, the policy can approximate the {LQR} solution and achieve a comparable performance. 
Although we successfully identified a valid Lyapunov function and a linear feedback controller, jointly optimizing these with stability and optimality guarantees remains challenging and is a key area for future research. Future work includes adaptation to nonlinear inverter dynamics, delay-aware controller design, decentralizing the controller to reduce communication needs, and improving robustness to parameter measurement errors and variability.

\bibliographystyle{IEEEtran}
\bibliography{reference}
\end{document}